\documentclass[aps,prb,superscriptaddress,reprint]{revtex4-1}
\usepackage{graphicx}
\usepackage{amsmath}
\usepackage{amssymb}
\usepackage{amsfonts}
\usepackage{filecontents}
\usepackage{color}
\usepackage[normalem]{ulem}
\usepackage{soul}
\usepackage{epstopdf}
\usepackage{hyperref}
\usepackage{fancyhdr}
\usepackage{bbold}
\definecolor{gold}{rgb}{0.85,.66,0}

\begin{document}

\title{Relaxation dynamics of spin 3/2 silicon vacancies in 4H-Silicon carbide}

\author{A. J. Ramsay}
\affiliation{Hitachi Cambridge Laboratory, Hitachi Europe Ltd., Cambridge CB3 0HE, United Kingdom}

\author{A. Rossi}
\affiliation{Department of Physics, SUPA, University of Strathclyde, Glasgow G4 0NG, United Kingdom}
\affiliation{National Physical Laboratory, Hampton Road, Teddington TW11 0LW, United Kingdom}

\date{\today}

\begin{abstract}{Room temperature optically detected magnetic resonance experiments on spin 3/2 Silicon vacancies in 4H-SiC are reported. The $m_s=+1/2\leftrightarrow -1/2$ transition is accessed using a two microwave frequency excitation protocol. The ratio of the Rabi frequencies of the $+3/2 \leftrightarrow +1/2$ and $+1/2\leftrightarrow -1/2$ transitions is measured to be $(0.901\pm 0.013)$. The deviation from $\sqrt{3}/2$ is attributed to small difference in g-factor for different magnetic dipole transitions. Whereas a spin-1/2 system is characterized by a single spin lifetime $T_1$, we  experimentally demonstrate that the spin 3/2 system has three distinct relaxation modes that can be preferentially excited and detected. The measured relaxation times are $(0.41\pm 0.02) T_{slow}=T_d= (3.3\pm 0.5)T_{fast} $. This differs from the values of $ T_p/3 =T_d= 2T_f $ expected for pure dipole ($T_p$), quadrupole ($T_d$), and octupole ($T_f$) relaxation modes, respectively, and implies admixing of the slow dipole and fast octupole relaxation modes.
 }
\end{abstract}

\maketitle


\section{Introduction}

The density matrix of a qubit is often decomposed into the identity and three Pauli spin-half matrices. The resulting Bloch-vector provides an intuitive picture of the spin-half dynamics, and the populations relax with a single spin-lifetime termed $T_1$.
A spin 3/2 system has four states, and is described by a 4x4 density-matrix. By extension, the relaxation of the four spin populations is described by three relaxation modes, characterized by three time constants. Furthermore, the 4x4 density matrix can be represented by  a multipole expansion of the identity, x3 dipole ($\mathcal{P}$), x5 quadrupole ($\mathcal{D}$), and x7 octupole ($\mathcal{F}$)modes providing a more intuitive representation of the spin-3/2 density-matrix. \cite{vanderMaarel_rev} This representation has advantages for understanding the spin-relaxation processes of S=3/2. For example, a dipole-like perturbation does not mix different order poles fixing the spin relaxation times such that $T_p/3=T_d=2T_f$, as recently discussed theoretically in the case of a fluctuating magnetic field acting on a silicon vacancy in SiC.\cite{Soltamov_Ncomm}$^,$\cite{Tarasenko_} However, as we will demonstrate, this is not the case in practice.

An accessible spin 3/2 system for testing this prediction is the V2 silicon vacancy in 4H-SiC \cite{Widmann_nmater,Carter_prb,Soltamov_Ncomm}. Recently, a number of groups have demonstrated that defects in SiC have optically accessible spins with coherence times on a par with diamond \cite{Koehl_nature,Carter_prb}. Unlike diamond, the manufacturing of SiC electronic devices is advanced. For example, n- and p-type doping can be routinely achieved and good quality $\mathrm{SiO_2}$ films can be deposited or grown on the surface, enabling CMOS processing. 
Due to the large breakdown voltages of SiC diodes and transistors, SiC devices are increasingly used in power electronic applications relevant to electric trains, cars and power transmission. As such, rapid improvement in materials and device quality is to be expected, and there is growing interest in SiC  for quantum devices \cite{Wolfowicz_ArXiv,Anderson_ArXiv,Calusine_apl,Lee_apl,Lohrman_Ncomm}.

Here, we report room temperature optically detected magnetic resonance (ODMR) experiments on an ensemble of  silicon vacancies in 4H-SiC. By using a two microwave frequency setup, the $+1/2\leftrightarrow -1/2$ transition can be detected optically \cite{Soltamov_Ncomm}$^,$\cite{Niethammer_prAppl}$^,$\cite{Nagy_Ncomm}. Rabi oscillations of all three magnetic dipole allowed transitions are measured. The ratio of the Rabi frequencies is compared to the value of $\sqrt{3}/2$, expected for $S_x$ matrix. A small difference in the in-plane g-factor for the $\pm 3/2\leftrightarrow\pm 1/2$ and $+1/2\leftrightarrow -1/2$ transitions is measured. In a typical $T_1$ measurement,\cite{Soltamov_Ncomm} a laser pulse initializes and detects the quadrupole state that decays exponentially  with a time constant of $T_d=131~\mathrm{\mu s}$. Here we use pulse sequences with two microwave frequencies that preferentially generate and detect the octupole and dipole states, and then measure their relaxation dynamics. The relaxation of the spin-3/2 is found to comprise of three modes with three time constants. The symmetry between exciting $\pm 3/2 \leftrightarrow \pm 1/2$ transitions implies that one of the relaxation modes is the quadrupole. However,
the fast relaxation mode decays much faster than expected for a pure octupole relaxation mode, with $T_{fast}<T_d/2$.
Therefore, contrary to expectations of ref. \cite{Soltamov_Ncomm}, the dipole and octupole relaxation modes are admixed due to a relatively fast relaxation between $+1/2\leftrightarrow -1/2$ states.

\section{Experimental details}

The silicon vacancy is  a point defect due to  a missing silicon atom. In 4H-SiC, there are two species of defect due to two inequivalent lattice sites with near hexagonal or cubic point symmetry \cite{Castelletto_rev,Heremans_ieee}. Here we nominally study an ensemble of V2 defects with a zero phonon line at 916 nm at low temperature, since it can be detected in ODMR at room temperature \cite{Carter_prb,Widmann_nmater}. The V2 is associated with the cubic k-site \cite{Ivady_prb}, and a zero-field fine structure splitting of 70 MHz. 

The sample used was purchased from CREE. The substrate is n-type LPBD. There is $30~\mathrm{\mu m}$ epilayer that is slightly n-type ($3\times 10^{15}~\mathrm{cm^{-3}}$). Due to the n-type substrate, the samples have a dark yellow color. 
The experiments were made using native defects of the epi-layer.

The  silicon vacancy is  optically polarized along the c-axis. To improve optical collection efficiency by a factor of 5-7, the chip is cleaved and mounted on its side to collect light perpendicular to the c-axis. \cite{Carter_prb} The 785nm pump laser is chopped with an acousto-optic modulator and coupled into a home-built microscope with a dichroic mirror. The laser (~14 mW) is focused on the side of the chip with a NA=0.75 air objective to a $\approx 1~\mathrm{\mu m}$ spot-size. The photoluminescence is fiber coupled to a Si-APD module. The count-rate is set to an optimum of about 3~MHz. Typically, $10^8$ counts are accumulated to achieve a standard deviation of $10^{-4}$. This results in long integration times. A typical trace in fig. 1 takes about 2hrs to measure, and fig. 3(d) took about 2 weeks.   The ac B-field is applied along the laser axis, perpendicular to the c-axis, using a loop antenna fashioned from a coaxial cable, and a dc B-field nominally along the c-axis is applied by positioning a permanent magnet. 


\section{Two-tone ODMR spectra}

\begin{figure}[ht!]
\begin{center}
\vspace{0.2 cm}
\includegraphics[width=1.0\columnwidth]{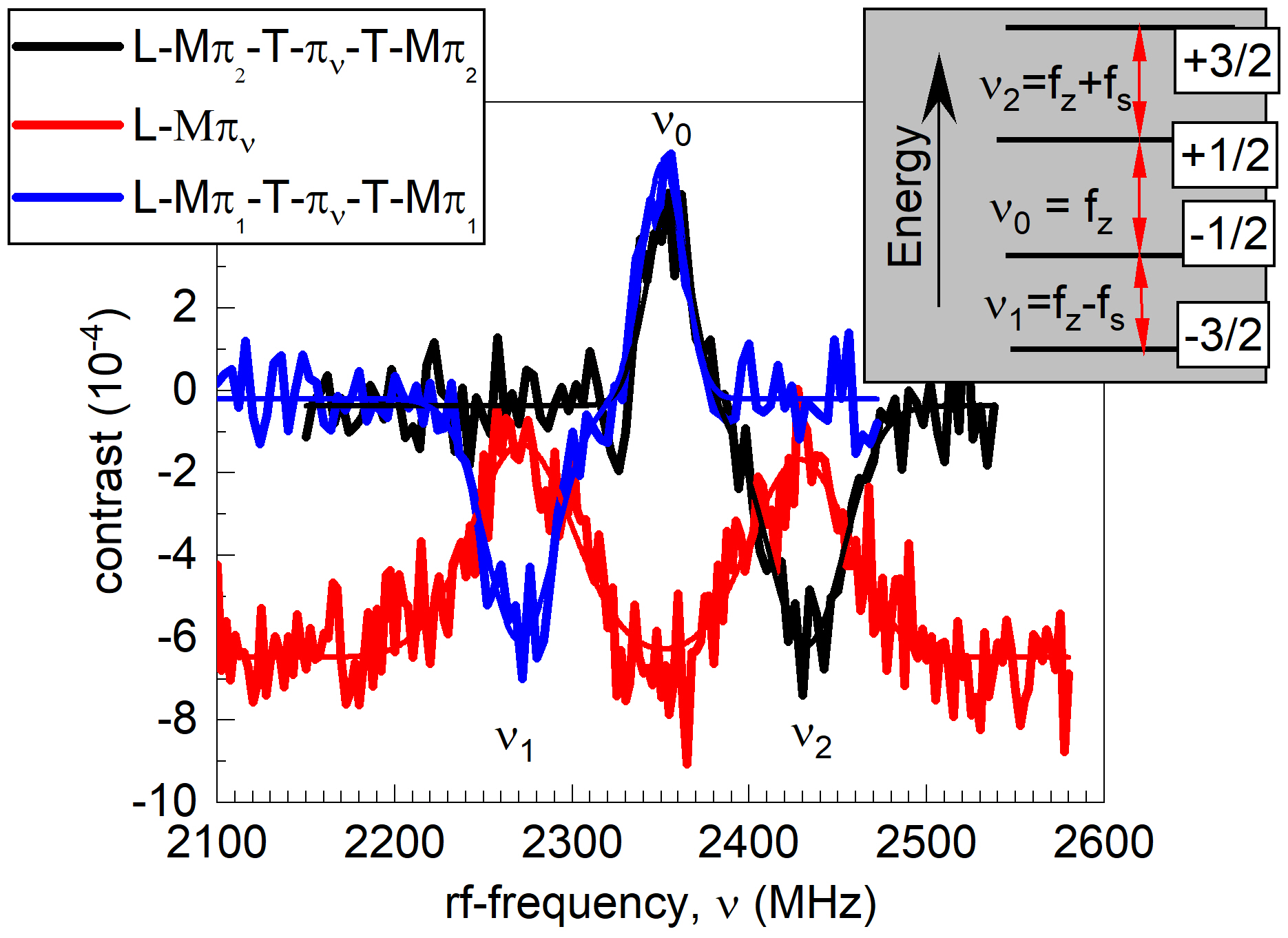}
\vspace{0.2 cm}
\end{center}
\caption{   (inset) Energy-level diagram of S=3/2 ground-state with dc B-field applied along c-axis. The magnetic dipole allowed transitions are labelled $\nu_{a}$. (main panel) (red) Single frequency ODMR spectra. ODMR is only sensitive to population difference between $\pm 3/2$ and $\pm 1/2$ states, hence the $\nu_0$ transition is not observed. (black/blue) Two frequency ODMR spectra, showing change in signal due to $\pi$-pulse inserted between two $\pi$-pulses tuned to $\nu_1$($\nu_2$) transitions. The $\nu_0$ transition is now observed. $2f_s=158\pm 4~\mathrm{MHz}$, $f_z=2350\pm 2 ~\mathrm{MHz}$. The offset is arbitrary.}\label{fig:2tonespec}
\end{figure}

The inset of fig. \ref{fig:2tonespec} shows the energy-level diagram of the spin-3/2 ground-state in a strong magnetic field of approximately 84 mT applied along the c-axis. To locate the transitions $\nu_1$ and $\nu_2$ a single frequency ODMR spectra is measured. This is done by applying a $\tau_L=3.8 ~\mathrm{\mu s}$ nonresonant laser pulse to generate a net spin in the $\pm 3/2$ states \cite{Carter_prb}. Then a 30-ns radio frequency (rf) $\pi$-pulse is applied, if this is resonant with either the $\nu_1$ or $\nu_2$ transition the $m_s=-1/2$ ($m_s=+1/2$) state is populated resulting in a slightly increased fluorescence when a second laser pulse is applied. A lock-in measurement comparing signals with and without the rf-pulse is used. Two peaks are observed. We note that the frequency splitting $2f_s=158\pm 4~\mathrm{MHz}$ is larger than the 140 MHz expected for $V2_{Si}^-$ defects in 4H-SiC. We do not attribute this to a misalignment of the magnetic field with respect to the c-axis since this would reduce the splitting. We note that in ref. \cite{Banks_prappl}, a slightly larger than expected splitting was also reported for single V2-related defect. Most likely, the larger than expected splitting here is due to an ensemble of silicon vacancies perturbed by nearby defects. There are a number of S=3/2 complexes associated with a negative silicon vacancy, with a nearby defect along the c-axis with various splittings \cite{Son2019}. In particular, the R2 complex is a close match with a splitting of 4D=157.6 MHz \cite{Son2019}.

To detect the $\nu_0$ transition, a two frequency pulse sequence $L-M\pi_{1,2}-T-\pi_{\nu}-T-M\pi_{1,2}$ is used. The notation summarizes the pulse-sequence with time going left to right. $L$ indicates the laser pulse for initialization and detection. $\pi_a$ indicates a $\pi$-pulse on the $a$-transition, $T$ a time-delay. For lock-in detection, the experiment alternates between two slightly different pulse sequences at half the repetition rate. $M$ precedes a pulse that is switched on and off at the repetition rate. Here the first $\pi_2$-pulse generates a population inversion between the $m_s=+1/2$ and $-1/2$ states, that can be driven by a $\pi_0$ pulse. The final $\pi_2$ pulse swaps the populations of the $+3/2$ and $+1/2$ states projecting a population between $\pm 1/2$ states into the measurement basis. The two frequency spectra are displayed in fig. \ref{fig:2tonespec}. An additional peak corresponding to $\nu_0$ is observed confirming that the ground state is spin 3/2. A dip at the $\nu_{1,2}$ transition is observed, since the lock-in compares the signals generated by sequences with three consecutive $\pi_{1,2}$ and one $\pi_{1,2}$ pulse, respectively. This is narrower than the single frequency peak, due to a spectral hole burning effect, whereby the pre-pulse selects a sub-set of defects to be probed by the $\pi_{\nu}$-pulse \cite{Soltamov_Ncomm}$^,$\cite{Niethammer_prAppl}. The $\nu_{2,1}$ peaks are absent since the $\pi_{1,2}$ and $\pi_{2,1}$ pulses interact with different spin-states, and the independent unmodulated signal generated by frequency scanned $\pi_{\nu}$-pulse is cancelled.

\section{Rabi frequencies of spin 3/2}

\begin{figure}[ht!]
\begin{center}
\vspace{0.2 cm}
\includegraphics[width=1.0\columnwidth]{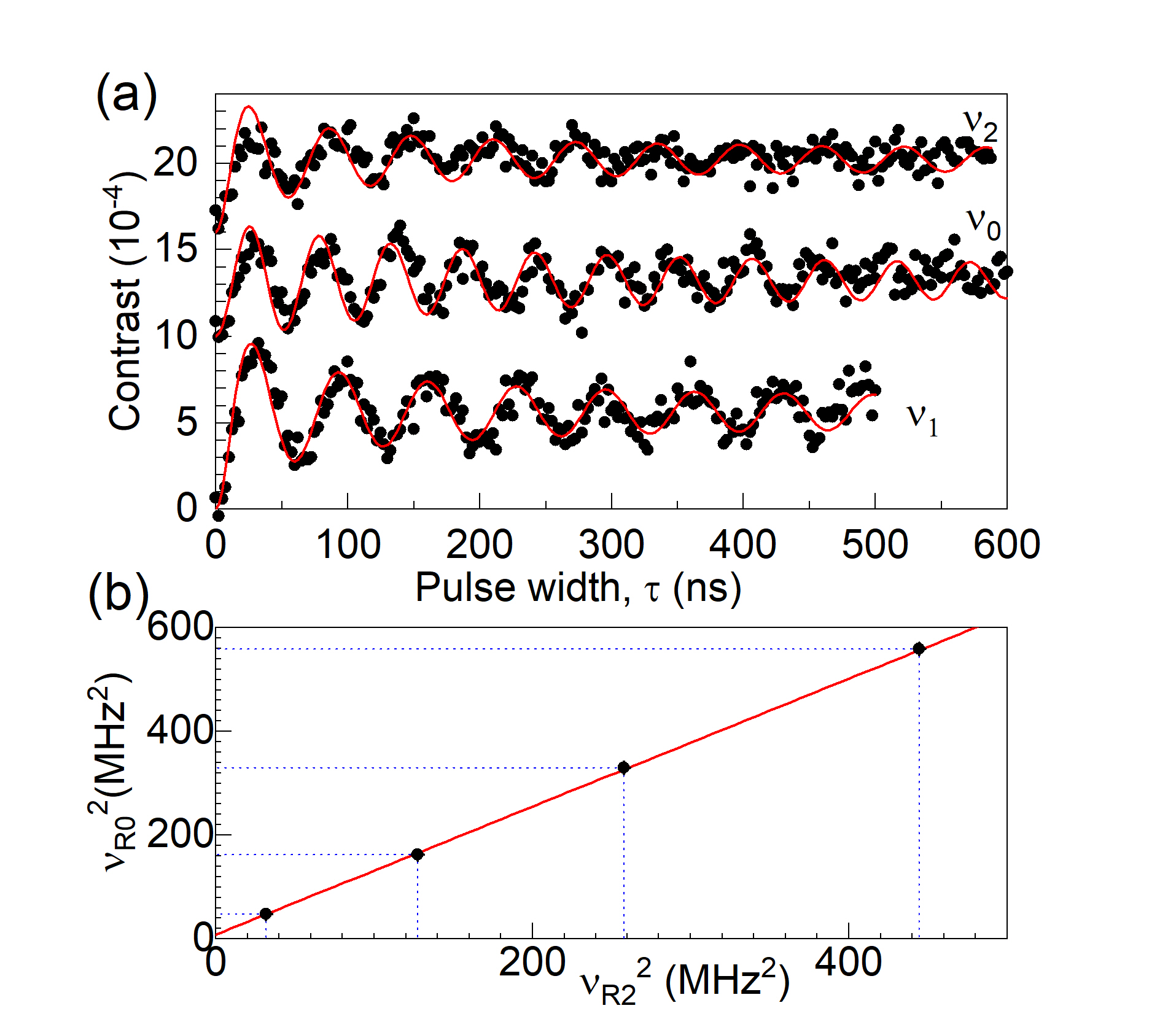}
\vspace{0.2 cm}
\end{center}
\caption{  (a) Rabi oscillations of all three magnetic dipole allowed transitions. (b) Plot of $\nu_{R0}^2$ versus $ \nu_{R2}^2$ Rabi frequencies squared, measured at rf-frequency $\nu_{rf}=2340.9~\mathrm{MHz}$ for various rf-powers.   Data offset for clarity. }\label{fig:fig1}
\end{figure}

A key property of spin-3/2 system is that the Rabi frequencies of the $\nu_{1,2}$ and $\nu_0$ transitions should have a ratio of $\sqrt{3}/2$ due to the ratio of the relevant elements of the S=3/2 $S_x$ matrix. Rabi oscillation measurements are made for all three transitions, see fig. 2. Since the pre-pulse selects a sub-set of defects, the damping of the $\nu_0$ Rabi oscillation due to ensemble broadening is reduced. To measure the ratio of the Rabi frequencies, Rabi oscillations are measured as a function of rf-power at a frequency of 2340.09 MHz by tuning the resonances with the external magnetic field. This eliminates changes in the applied power due to the frequency response of the loop-antenna. The Rabi frequencies are extracted by fit to a model that accounts for the ensemble broadening, the red-lines in fig. 2(a) give example fits. Figure 2(b) displays a plot of Rabi-frequency squared of the $\nu_2$ and $\nu_0$ transitions. This plot eliminates effects of saturation of the power applied at the loop antenna, and the intercept accounts for contributions to the effective Rabi frequency due to small detunings, or dephasing. The gradient gives $R^{-2}$, where $R$ is the ratio of the $\nu_{R2}/\nu_{R0}$ Rabi frequencies, with $R=0.901_{-0.013}^{+0.009} $. This is larger than the value of $\sqrt{3}/{2}$ expected for an isotropic S=3/2 system.

The deviation in the ratio $R$ from $\sqrt{3}/2$ can be interpreted as a slight difference in the g-factors of the two transitions. We define an anisotropic g-factor tensor $g_{ij,k}$, such that the Zeeman Hamiltonian  is $H_Z=\mu_B g_{ij,k}S_{ij,k}B_k$. The in-plane ac B-field is aligned along $x$, and $R=\frac{\sqrt{3}}{2}\frac{g_{+3/2\leftrightarrow 1/2, x}}{g_{+1/2\leftrightarrow-1/2,x}}$. Then following notation of ref. \cite{Simin_PRX}, we further define a deviation from $g_{ij,x}=g_{\perp}$, such that $g_{+3/2\leftrightarrow+1/2,x}=g_{\perp}+g_{2\perp}$, and $g_{+1/2\leftrightarrow -1/2,x}=g_{\perp}-g_{2\perp}$, and deduce $\frac{g_{2\perp}}{g_{\perp}}=+0.019^{+0.005}_{-0.007}$. This is consistent with value of $\frac{g_{2\perp}}{g_{\perp}} = 0.0\pm 0.05$ reported in ref. \cite{Simin_PRX}. For further details of the analysis, see appendix A.

\section{Spin relaxation modes}

\begin{figure}[ht!]
\begin{center}
\vspace{0.2 cm}
\includegraphics[width=1.0\columnwidth]{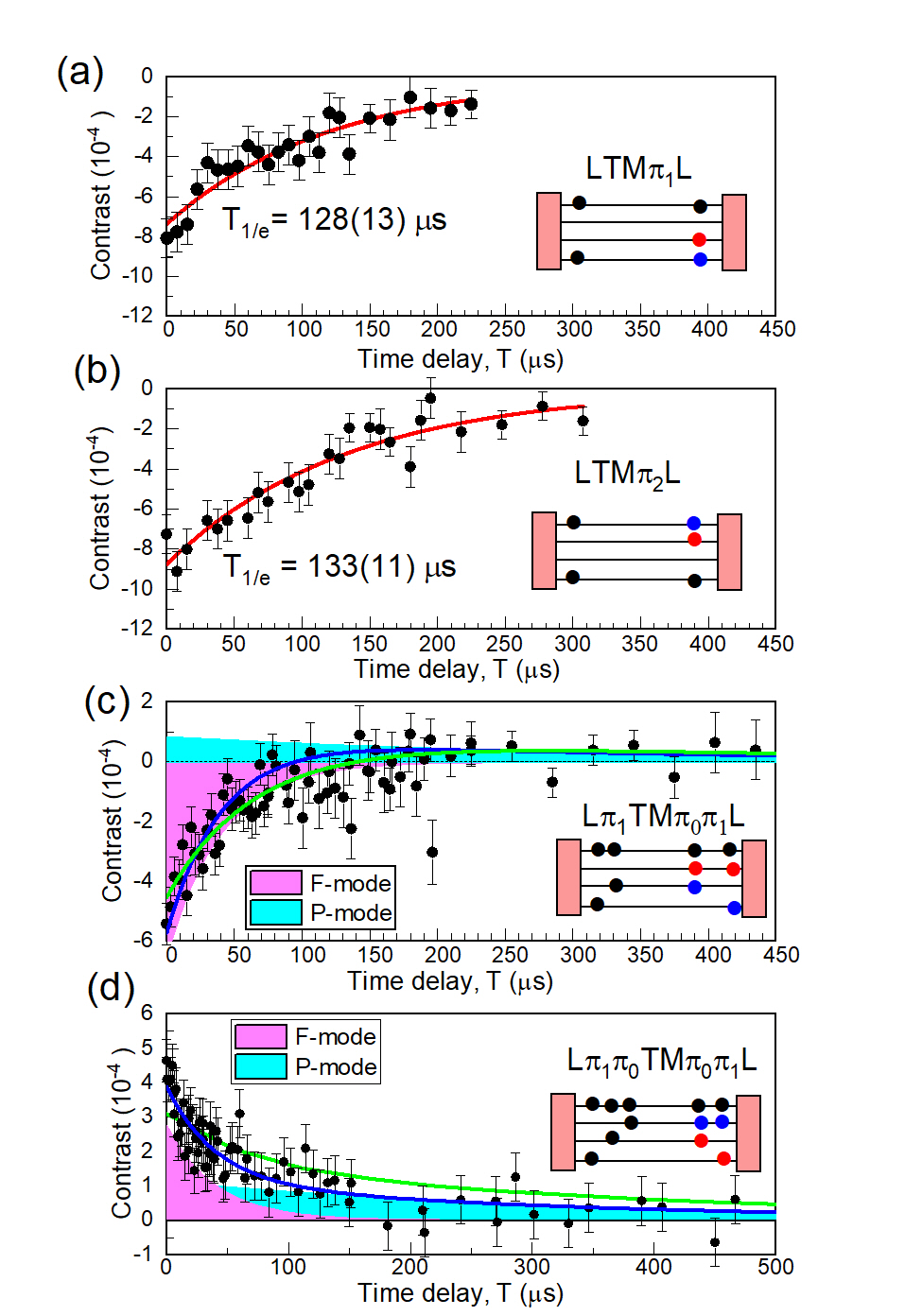}
\vspace{0.2 cm}
\end{center}
\caption{  Spin relaxation of qudit modes. Error bars are estimated from total counts, $N$, as error$=1/\sqrt{N}$.(a,b) Pure quadrupole relaxation exciting $\nu_1$ and $\nu_2$ transitions, respectively. (c) Mostly Octupole relaxation (d) Mixture of octupole and dipole modes.   (green) Fit with $\gamma_2=2\gamma/3$, where $T_p/3=T_d=2T_f=131~\mathrm{\mu s}$.  (blue) Fit to (d) yields $\gamma_2/\gamma=1.33\pm 0.27$. In (c), blue line is a calculation using $\gamma_2/\gamma=1.33 $. (magenta/cyan) Shows the decomposition of the spin relaxation into the fast and slow decay modes.  The signal offset is zero. Insets: The cartoons illustrate the pulse sequence. The 4 lines represent the spin states in order of energy, and the pink blocks the laser pulse. Black, red, and blue circles represent populations that are constant, and only present in case of modulated pulse ($M\pi_i$) on, and off, respectively. The cartoon depicts the case of no spin relaxation, and ideal $\pi$-pulses.   }\label{fig:fig3}
\end{figure}

The spin-relaxation dynamics of the 4-state S=3/2 system can be described by the matrix $\mathcal{R}$, such that $\dot{\rho}_{ii}=-\mathcal{R}_{ij}\rho_{jj}$, where the diagonal of the density-matrix $\rho_{jj}$ represents the populations in the $m_s=(+3/2,+1/2,-1/2,-3/2)$ states respectively. If we assume that only $\Delta m_s=\pm 1$ transitions are allowed, the relaxation matrix has the form
\begin{eqnarray}
\mathcal{R}=\left(
  \begin{array}{cccc}
    \gamma_1 & -\gamma_1 & 0 & 0 \\
    -\gamma & \gamma_1+\gamma_2 & -\gamma_2 & 0 \\
    0 & -\gamma_2 & \gamma_3+\gamma_2 & -\gamma_3 \\
    0 & 0 & -\gamma_3 & \gamma_3 \\
  \end{array}
\right).
\end{eqnarray}
Therefore the spin-relaxation dynamics can be decomposed into 4 eigen-modes. The first is the identity ($\rho=\mathbb{1}$) with a decay-rate of zero. In the case where $\gamma_1=\gamma_3=\gamma/2$, the eigen-values of $\mathcal{R}$ are:
\begin{eqnarray}
\frac{1}{T_d}=\gamma, \nonumber \\
\frac{1}{T_{fast,slow}}=\frac{\gamma}{2}+\gamma_2\pm\frac{1}{2}\sqrt{\gamma^2+4\gamma_2^2}
\end{eqnarray}
In the special case considered in ref. \cite{Soltamov_Ncomm} where an effective magnetic field gives rise to fluctuations with dipole-like symmetry, $\gamma_2=2\gamma/3$, and the eigenmodes are the pure dipole ($\mathcal{P}_0=(3,1,-1,-3)/\sqrt{20}$), quadrupole ($\mathcal{D}_0=(1,-1,-1,1)/2$), and octupole ($\mathcal{F}_0=(1,-3,3,-1)/\sqrt{20}$) modes with relaxation times:  $T_p=3T_d=6T_f$. If however, the fluctuations have a component with quadrupole-like symmetry, then $\gamma_2 \neq 2\gamma/3$, and admixing of the dipole and octupole relaxation modes occurs.

To test this picture, a series of measurements are made to preferentially generate and detect the dipole, quadrupole and octupole modes, and measure their relaxation rates. Figure \ref{fig:fig3}(a) shows the relaxation of a pure quadrupole mode using the pulse sequence ($L-T-M\pi_1-L$) to initialize in $\mathcal{D}_0$, and then make a projective measurement at time-delay $T$. This is the usual protocol used to measure the `$T_1$' time \cite{Soltamov_Ncomm}$^,$\cite{Kasper_ArXiv}. Consistent with previous reports, a single exponential decay is observed. A similar measurement is made on the $\nu_2$ transition, see \ref{fig:fig3}(b). Decay times of $T_{d\nu_1}=128\pm 13~\mathrm{\mu s}$, and $T_{d\nu_2}=133 \pm 11~\mathrm{\mu s}$ are measured, these are the same within experimental error. This implies a symmetry between the $\nu_1$ and $\nu_2$ transitions, such that $\gamma_1\approx \gamma_3$, and confirms that the quadrupole is an eigenmode of the relaxation matrix $\mathcal{R}$, with $T_d=131\pm 8~\mathrm{\mu s}$.
 This is towards the short end of the $T_1$ values reported for $V2_{Si}^-$ in natural 4H-SiC. Together with the relatively short spin-echo times, $T_2(SE)<2~\mathrm{\mu s}$ this suggests the defect density of the sample is relatively high $N_V\sim 10^{16}~\mathrm{cm^{-3}}$ in this sample \cite{Kasper_ArXiv}$^,$\cite{Brereton_ArXiv}, or the relatively high background n-type doping ($n=3\times 10^{15}\mathrm{cm^{-3}}$) may reduce the stability of the defects charge state.

If the dipole and octupole relaxation modes are uncoupled, the pulse sequences ($L-\pi_1-T-(M\pi_0)\pi_1-L$), and ($L-\pi_1\pi_0-T-(M\pi_0)\pi_1-L$), as used for figs. 3(c,d), should generate a mostly octupole signal $S(T)\propto +0.2e^{-T/T_p} - 1.2e^{-T/T_f}$, and a mixed $S(T)\propto +0.5e^{-T/T_p} + 0.5e^{-T/T_f}$, respectively. This is independent of the fidelities of the $\pi$-pulses, which only affect the overall signal amplitude. These sequences are designed to cancel the quadrupole contribution, regardless of dipole/octupole admixing, and the fidelities of the $\pi$-pulses.

Figure \ref{fig:fig3}(c) shows the relaxation of the mostly octupole mode using the pulse sequence ($L-\pi_1-T-(M\pi_0)\pi_1-L$). The offset is subtracted using a double lock-in method (see appendix B), and since the signal crosses zero there are two relaxation components with opposite sign, as expected.
 A single exponential fit yields $T_{1/e}=48\pm 7~\mathrm{\mu s} \approx T_f'$, this is noticeably faster than the expected value of $T_f=T_d/2=66~\mathrm{\mu s}$. Figure \ref{fig:fig3}(d) shows the relaxation of a mixture of  the octupole and dipole modes using the pulse sequence ($L-\pi_1\pi_0-T-(M\pi_0)\pi_1-L$). After a fast initial decay, a long tail is observed, but the ratio of the fast to slow components is larger than the expected ratio of 1:1, further demonstrating admixing of the dipole and octupole relaxation modes.

The data is modelled using Eq. (1). The initial state after laser initialization is a pure quadrupole state. The $\pi$-pulses are modelled as matrix operations with a fidelity $F_0=0.84$, and $F_{1,2}=0.74$ to account for imperfect inversion, and are measured by comparing to Rabi oscillation data, assuming the long time limit in the signal matches the inversion point. We note that the fidelities have a weak effect on the ratio of relaxation modes observed in fig. 3(d) only. Otherwise the only effect is on the overall amplitude of the signal. Following initialization of the spin, the populations-vector is decomposed into the relaxation eigen-modes, and following a time-delay $T$, the projection of the read-out sequence onto quadrupole measurement basis is calculated.

To test if the relationship $T_p/3=T_d=2T_f$ holds, the data in figs. 3(c) and (d), are fitted to the model of Eq.(1) with fixed $\gamma_2=2\gamma/3$, and $T_d=131~\mathrm{\mu s}$ as determined from fit to figs. 3(a) and (b), see green lines of figs. 3(c,d). The only fitting parameter is the amplitude. A clear deviation from the constrained model is found for fig. 3(d), where the mixture of the P and F relaxation modes probed in the measurement are sensitive to $\gamma_2/\gamma$. The data of fig. 3(d) is then fitted to model, with $\gamma_2$ as a fitting parameter yielding $\gamma_2=(1.33\pm 0.27 )\gamma$, where the error is given by the 95\% confidence level. To check consistency of the model, we then fit fig. 3(c) with fixed $\gamma_2=1.33\gamma$, shown as a blue line. We then infer   $T_d/T_{fast}=3.3\pm 0.5 \neq 2$, and $T_d/T_{slow}=0.41^{+0.01}_{-0.02}\neq 1/3$ for the relaxation rates, with $T_{fast}=40\pm 6 ~\mathrm{\mu s}$, and $T_{slow}=320_{-22}^{+26}~\mathrm{\mu s}$.

\section{Conclusions}

 To conclude, we have presented two microwave tone optically detected magnetic resonance experiments on an ensemble of  silicon vacancies in 4H-SiC, with spin 3/2. These measurements provide access to all the magnetic-dipole allowed transitions. A comparison of the Rabi frequencies for the $\pm 3/2\leftrightarrow \pm 1/2$, and $+1/2\leftrightarrow -1/2$ transitions allows us to measure a slightly different in-plane g-factor for these transitions. The relaxation of the spin 3/2 system is shown experimentally to have three relaxation modes that can be preferentially generated and detected by choosing a particular microwave pulse sequence. This contrasts with a spin-1/2 system characterized by a single $T_1$-time. The spin-relaxation is approximately symmetric with respect to interchange of the $\pm 3/2\leftrightarrow \pm 1/2$ transitions, indicating  a pure quadrupole relaxation mode. Contrary to theory in ref. \cite{Soltamov_Ncomm}, the decay of the short-lived octupole-like mode is faster than expected for a fluctuating in-plane B-field. This indicates mixing of the octupole and dipole relaxation modes, since a perturbation with dipole symmetry cannot mix different order poles. This suggests an additional fluctuation with quadrupole symmetry that mixes the dipole and octupole modes of odd order. \cite{vanderMaarel_rev}
 This may be the result of dipolar interactions with neighboring electron spin-1/2 defects, where the energy-cost of flipping a parasitic spin-1/2 matches the Zeeman splitting of the $+1/2\leftrightarrow-1/2$ transition, and where the $\pm 3/2\leftrightarrow \pm 1/2$ transitions are protected by an energy mismatch due to the crystal-splitting $2D\approx 70~\mathrm{MHz}$. Or may arise due to fluctuations in the crystal-field splitting $D$. This work demonstrates that $T_1=T_d$ measurements do not provide complete information on spin-relaxation dynamics of spin 3/2 systems.

\acknowledgements

 We thank the following people for their help. J. A. Haigh and R. A. Chakalov for technical assistance. Akio Shima, Kumiko Konishi and Keisuke Kobayashi of Hitachi CRL for donating the material; R. Webb of EPSRC ion implantation facility at University of Surrey for C-ion implantation. G. W. Morley, H. Knowles, B. Pingault, and D. Kara for advice on ODMR measurements. A. R. acknowledges financial support from UKRI Industrial Strategy Challenge Fund through a Measurement Fellowship at the National Physical Laboratory.

\appendix

\section{Analysis of Rabi oscillation ratio}

 Because the Rabi frequency is much smaller than the splitting, $\nu_R<25~\mathrm{MHz} \ll 2f_s=158~\mathrm{MHz}$, and the inhomogeneous broadening dominates the damping, $\Gamma^* \gg 1/T_1, 1/T_2$ we treat the system as an ensemble of detuned ideal two-level systems. The effective Rabi frequency of the transitions is \cite{2LStext}
\begin{eqnarray}
\nu_{R0}^2 = (g_{+1/2\leftrightarrow -1/2,x}\mu_B B_{ac}(P_{rf})) ^2 + \delta_0^2, \\
\nu_{R2}^2=(g_{+3/2\leftrightarrow +1/2,x}\frac{\sqrt{3}}{2}\mu_BB_{ac}(P_{rf})^2 +\delta_2^2
\end{eqnarray}
where $\delta_{i}$ accounts for an error in the detuning between the rf-drive, and the transition-i, and a tiny shift due to intrinsic dephasing. $g_{ij,k}$ is the g-factor tensor, such that the Zeeman-term in the Hamiltonian is $H_Z=\mu_B g_{ij,k}S_{ij,k}B_k$, where $S_{ij,k}$ is the ij-element of $S_k$ spin-3/2 matrix. At low rf-powers ($P_{rf}$), the ac B-field $B_{rf}\propto \sqrt{P_{rf}}$. Eliminating the unknown $\mu_BB_{ac}(P_{rf})$ yields,
\begin{eqnarray}
\nu_{R0}^2= \frac{2g^2_{+1/2\leftrightarrow -1/2}}{3g^2_{+3/2\leftrightarrow +1/2}}\nu^2_{R2} + constant.
\end{eqnarray}
The red-lines in fig. 2(a) show example fits used to extract the Rabi frequencies. The Rabi oscillation signal $S(T)$  an ideal two-level system with detuning, as given by Eq. (3.16) of ref. \cite{2LStext}, is averaged over a Gaussian distribution of detunings, $\Delta$.
\begin{eqnarray}
S(T)\propto \int d\Delta \frac{\Omega^2_R}{\Lambda^2_R}\sin^2{(\frac{\Lambda_R T}{2})} e^{-\frac{\Delta^2}{\Delta_0^2}}
\end{eqnarray}
where $\Omega_R=2\pi\nu_R$ is the Rabi frequency and $\Lambda_R^2=\Omega^2_R+\Delta^2$ is the effective Rabi frequency.
The model has three fitting parameters: the amplitude, the inhomogeneous broadening $\Delta_0$, and the Rabi frequency $\nu_{Ri}$. The gradient of fig. 2(b), gives the ratio of the $R=\nu_{R2}/\nu_{R0}=0.901\pm 0.007> \sqrt{3}/2$.

To evaluate systematic errors, the ratio R was calculated as a function of B-field angle, inhomogeneous broadening and E-parameter. Inhomogeneous broadening effectively dresses the Rabi frequency increasing the ratio R by $\Delta R_{inhomo}\ll +0.004$. The effects of misaligned B-field, and strain are computed by considering the zero B-field Hamiltonian $H_0= D(S^2_z-5/4)+E(S_x^2-S_y^2)$, where $D=35 ~\mathrm{MHz}$, and $E$ is expected to be small \cite{Widmann_nmater}, and an isotropic Zeeman-Hamiltonian, where $g_{ijk}=g$. The ratio $R$ increases with out-of-plane B-field, with a maximum value of R=0.90 at $90^{\circ}$. A large misalignment angle of $10^{\circ}$ is found to increase $R$ by $\Delta R=+0.0009$. We find that $\Delta R_E= -3.5\times 10^{-4} ~\mathrm{MHz^{-1}}E$. An upper limit of $\vert E\vert < 18~\mathrm{MHz}$ is given by the splitting between the $\nu_1$ and $\nu_2$ transitions, with $2f_s=4\sqrt{D^2+E^2}$, \cite{Widmann_nmater}, yielding $\vert \Delta R_E \vert < 0.006$. Combining these errors yields $R=0.901^{+0.009}_{-0.013}$.

\section{Double-lock-in method}

The data collected in sec. V uses a double lock-in to achieve a stable zero offset. We use a gated  APD module with 2.5 dark cps (Laser Components Count-10). The 15-ns TTL output is switched between two channels of an open-source photon counter \cite{FPGA} using a microwave-switch (Mini-circuits ZWASWA-2-50DRA+). The switch slightly attenuates the TTL pulse, and it is necessary to terminate the FPGA inputs with $100\Omega$, rather than the usual $50\Omega$, to get reliable counting. An external 100-MHz clock is used (AEL9700CS), since the internal 48MHz clock is too slow.

In general, the rf-pulse sequence alternates between sequences S1 and S2 which are directed to channels 1 and 2. At a slower frequency, typically $\sim 0.1 \mathrm{Hz}$ the order of the sequences is swapped so that S2 and S1 are directed to channels 1 and 2. By calculating $S= \frac{S1_1-S2_2}{S1_1+S2_2} -\frac{S2_1-S1_2}{S2_1+S1_2}$, a small  imbalance in the detection channels $\sim10^{-4}$ is cancelled.

\end{document}